\documentclass[preprint,12pt]{elsarticle}
\usepackage{epstopdf}
\usepackage{amssymb}

\journal{Journal of Neuroscience Methods}

\begin{document}

\begin{frontmatter}

\title{A Wavelet Based Algorithm for the Identification of Oscillatory Event-Related Potential Components}

\author[aka]{Arun Kumar A. \corref{cor1}} \ead{aka.bhagya@gmail.com}
\cortext[cor1]{Corresponding author. Tel.: +91 9446186842}
\author[aka]{Ninan Sajeeth Philip} \ead{nspp@iucaa.ernet.in}
\address[aka]{St. Thomas College, Kozhencherry 689641, Kerala, India} 
\author[vjs]{Vincent J. Samar}\ead{vjsncr@rit.edu}
\address[vjs]{Rochester Institute of Technology, 52 Lomb Memorial Drive, Rochester, NY 14623, USA}
\author[sjs]{James A. Desjardins} \ead{jdesjardins@brocku.ca}
\author[sjs]{Sidney J. Segalowitz} \ead{ssegalowitz@brocku.ca}
\address[sjs]{Brock University, 500 Glenridge Ave., St. Catharines, ON L2S 3A1, Canada}

\begin{abstract}
Event Related Potentials (ERPs)  are very feeble alterations in the  ongoing Electroencephalogram (EEG)  and their detection  is a challenging problem. Based on the unique time-based parameters derived from wavelet coefficients and the asymmetry property of wavelets a novel algorithm to separate ERP components in single-trial EEG data is described. Though illustrated as a specific application to N170 ERP detection, the algorithm is a generalized approach that can be easily adapted to isolate different kinds of ERP components. The algorithm detected the  N170 ERP  component with a high level of accuracy. We demonstrate that the asymmetry method is more accurate than the matching wavelet algorithm and t-CWT method by 48.67  and 8.03 percent respectively. This paper provides an off-line demonstration of the algorithm and considers issues related to the extension of the algorithm to real-time  applications.

\end{abstract}


\end{frontmatter}

\section{Introduction}\label{intro}
The ability to detect single-trial event related potentials (ERPs) in real-time EEG signals has many clinical and research applications, particularly in the field of brain computer interfaces \citep{barret,birbaumer}.  
ERPs appear as dynamic alterations in ongoing EEG frequency components that are very feeble signals compared to background EEG and have low signal-to-noise ratio when recorded from electrodes attached to the scalp. Detection of ERP components from real-time EEG is therefore a challenging problem.

Studies based on unique ERP features \citep{robert} and conventional anomaly detection algorithms like filtering \citep{cong} and matching pursuit have given limited success for the detection of ERPs \citep{tomas}. However, use of a band-pass filter specifically tuned for EEG/ERP frequencies improves signal to noise ratio. While  \cite{watun} have shown that linear FIR filters may not perform optimally, filters tailor made for the particular ERP, such as Woody filters and  matched filters provide better detection accuracy \citep{ford,serby}. Also \cite{jind} have shown that the Hilbert-Huang transform \citep{hhtr} could further improve the discrimination power of filter based approaches. But these filtering variations are limited in their success rate with real time signal detection.

The joint time-frequency analysis of signals is a solution to overcome the limitations of conventional filtering techniques. Wavelets can be used for the joint time-frequency analysis of EEG signals and they provide more robust measures for the detection and analysis of ERP components \citep{blanco}.

\cite{vince} and \cite{quian} have presented evidence that wavelets may improve the extraction and analysis of ERP waveforms. The applications of wavelets to ERPs are broad ranging, including joint time-frequency analysis of ERPs \citep{vince1}, artifact removal \citep{joe} and event detection \citep{demir,vince3}. Furthermore, features derived from wavelet coefficients \citep{anna,trejo}  perform well in preprocessing \citep{kala} stages of classification problems using statistical learning algorithms \citep{vahid,browne}.

A variety of wavelet based methods have been used to study different aspects of EEG/ERP signals. Methods based on the discrete wavelet transform (DWT)\citep{burger,herrera} and continuous wavelet transform(CWT) \citep{bostanov1}, in combination with other statistical measures \citep{lim}, have been tried for the detection as well as the analysis of ERPs. Variable threshold schemes \citep{fatour} and wavelet packet analysis \citep{bern} have shown reasonable results when there is no background noise. A more accurate method proposed by \cite{chapathesis,chapa} for the detection and multiresolution analysis of ERPs is computationally expensive and at times gave unbounded errors.

We propose a generalized, yet simple and powerful scheme using wavelets to detect specific ERP components from EEG data. The algorithm  makes use of a less-used asymmetry property of wavelets along with time base properties of the target ERP component.  Asymmetry is associated with the  observed phase shift of wavelet coefficients when the wavelet transform is performed on a signal. The amount of phase shift produced by a wavelet for a specific ERP is unique and can be used for its detection. A detailed description of wavelet asymmetry is given in section \ref{asymmetry}.

Our method can be used to detect different ERP components by changing the wavelet basis function and time base parameters. It is also automated with self-correction and validation mechanisms using wavelet features and time-based properties of the ERP components as a guide. This paper outlines the general approach, provides an off-line demonstration of the detection performance of the algorithm using the N170 component of ERPs to 'face' versus 'non-face' stimuli, and considers issues related to the extension of the algorithm to real-time applications.

\section{Asymmetry in wavelets}\label{asymmetry}
When the phase difference between the input and output signal is zero, the corresponding digital filter is said to be linear. Linear filters are symmetric.
Wavelets behave exactly like filters and a wavelet function $\psi$ when convolved with an input signal $\mathit{f(t)}$ will project the signal onto an orthogonal subspace $\xi$ as $\hat{f}(\xi)$. In terms of symmetry, a wavelet filter with coefficients $a_n$ is linear if the phase of the function $\mathit{a(\xi) = \sum_n a_n e^{i n \xi}}$ is a linear function of $\xi$ for some $\mathit{l \in \mathbb{Z}} $ \citep{daube}. This essentially means that the filter delays each frequency in the input signal in equal amounts at the output. The phase delay and group delay of such filters will have a flat profile for all the input frequencies similar to the one shown in Figure 1a and 1b .

\begin{figure}[!h]
\centering
\includegraphics{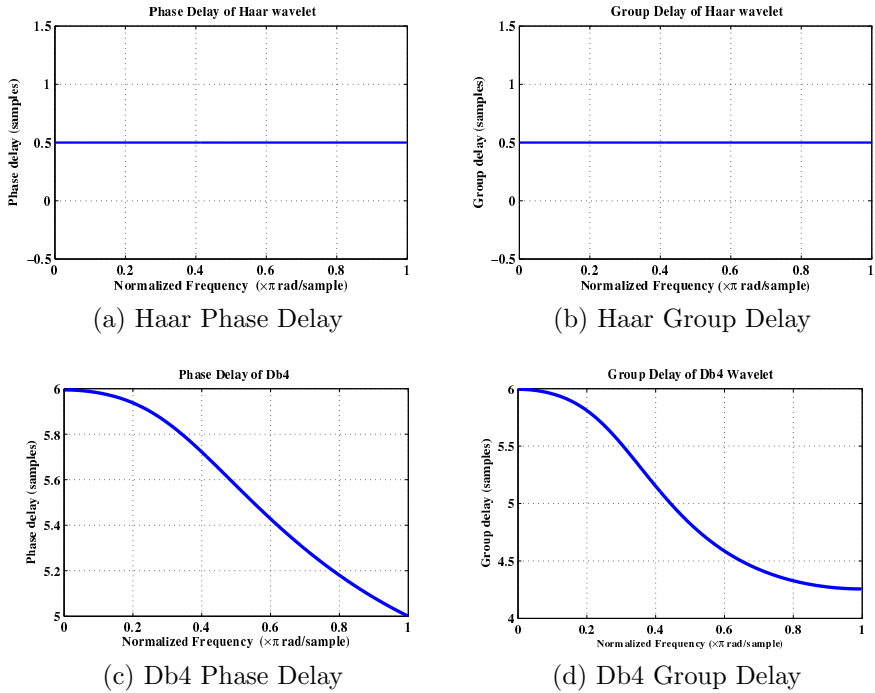}
\caption{(a) and (b) shows the flat response of the Haar wavelet which is a symmetric filter. Db4 is an asymmetric filter whose phase delay and group delay are shown by (c) and (d). Since the wavelet filter is not symmetric, the filter responds with time delays.}
\end{figure}\label{fig1}

When the filter response is non linear, which means different frequencies are shifted by different amounts at the output, the filter is said to be asymmetric. The phase delay response and group delay response of asymmetric filters will not have a flat profile. An example is shown in Figure 1c and 1d  where each frequency component of the signal is shifted by a specific number of samples. 

\subsection{Asymmetry as a measure}

Filter phase response is quantified in terms of group delay $\tau(\omega)$ which is given as
\begin{equation}
\tau(\omega) = - \frac{d\theta(\omega)}{d\omega} \label{eq2}
\end{equation}
where $\theta(\omega)$ is the phase of the wavelet filter $\mathit{H(\omega)}$.

Let $\delta_{k}$ be the phase shift introduced by an asymmetric wavelet to a signal $\mathit{f(t)}$ at the analysis scale $\phi_{k}$. $t_{k}$ is the corresponding shift on the time scale due to $\delta_{k}$. Both $\delta_{k}$ and $t_{k}$ are  due to group delay $\tau(\omega)$. When ERPs are convolved with asymmetric wavelet function $\psi$  the output phase shift $\delta_{k}$ and the corresponding shift on timescale $\mathit{t_{k}}$ will stay bounded within a unique value. This is because ERPs are band limited signal components and the corresponding wavelet coefficients generated will be shifted with respect to the phase response of the wavelet at those frequencies. 

Therefore the value of $\mathit{t_{k}}$ will be unique for a specific ERP when it is convolved with the best match of its wavelet function. When there is maximum resemblance of the wavelet with the signal under analysis, the wavelet coefficient generated will be maximum. This is because wavelet is a function that depends on the spectrum  of the signal and not its amplitude. 

In our method we combine the unique phase shift due to asymmetry and corresponding wavelet coefficients to detect ERP components from single trial EEG data. The only requirement of deriving the asymmetry measure is that the ERP should be oscillatory like a wavelet, so that the best fit wavelet does not have energy exceeding the non oscillatory components of the ERP. For ERP components that are not oscillatory, such as slow wave shifts, the group delay induced by the wavelet filter will be very minimal because there maybe no frequency components in the group delay band of the wavelet.  

\section{Application of Wavelet Asymmetry to detect N170 ERP Component}
To demonstrate the feasibility of the asymmetry measure as a detection metric, we have chosen the N170 ERP as an example. The N170 is an ERP which is oscillatory and has possible wavelet candidate matches.

\begin{figure}[!h]
\begin{center}
\includegraphics{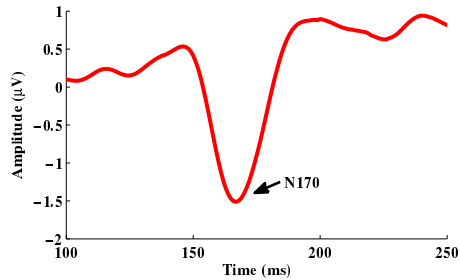}
\caption{The typical N170 ERP has a negative dip that depends on the nature of the stimuli shown to the subject.}\label{fig2}
\end{center}
\end{figure}

Currently, face recognition is one of the challenging and active research areas of cognitive science.  The N170 ERP is a characteristic response to the presentation of a face stimulus and is used to illustrate the implementation of our algorithm. A face stimulus presented to a human observer elicits a large negative amplitude between two positive going peaks in the EEG between 130ms and 200ms following the stimulus as shown in Figure \ref{fig2}. This large negative amplitude N1 which is observed at the occipito-temporal sites is termed the N170 \citep{bentin}.

\subsection{Data Acquisition}
The EEG data used in this study came from the study by \cite{james}. Ten healthy right handed adults in the age range 22-37 (9 females) were presented with 26 images. The images were gray scale front views of 16 faces, 8 houses and two half-circle checkerboards. The face stimuli were of four identities with emotions of anger and fear either in upright or inverted orientations. The house stimuli comprised four identities which were also either upright or inverted. The stimulus samples can be seen in \citep{james}. The stimuli were presented to the subjects in a dark room on a CRT monitor at 1024$\times$768 resolution with a refresh rate of 60 Hz, with the images on a black background. The images were constantly shown for the duration of the task, where the subjects were to respond with both hands on a four-key keypad. There were 8 trial blocks with 300 trials in each block. 

In addition we used 400 samples of resting EEG collected by \cite{james} while subjects passively watched the Windows XP star pattern screen saver. Each segment was 1 minute long.

The EEG data was recorded with 128 channels using the BioSemi ActiveTwo system at a sampling rate of 512 Hz. Artifacts and noise were removed from the EEG data using advanced data preprocessing techniques including Independent Component Analysis as described in \cite{james}.

\section{Methods}

\subsection{Selection of Wavelet \& Scale for N170 ERP} \label{scalesel}
The wavelet basis function and the scale are critical factors for any kind of component identification based on the CWT \citep{vince3}. Previous studies   have shown various methods for selecting the best wavelet basis \citep{brechet,flanders,niel}.

We have used a generalized algorithm for the selection of a wavelet basis function and scale which is illustrated by  \cite{rafee}. In this method, a wavelet basis function is chosen such that maximum wavelet coefficients are obtained when there is maximum resemblance between the wavelet and the signal in terms of its spectral features. 

For example, consider the continuous wavelet transform of a signal $\mathit{x(t)}$ as :

\begin{equation}
C (a,b) = \int_{R} x(t) \psi_{a,b} \mathrm{dt} = \int_{R} x(t) \frac{1}{\sqrt{a}} \psi \left( \frac{t -b}{a} \right) \mathrm{dt}  \label{eq4}
\end{equation}

where $\mathit{C (a,b)}$ are the wavelet coefficients, $\psi(\cdot)$ the wavelet function, $\mathit{a}$ the scaling factor and $\mathit{b}$ the shifting parameter. The continuous wavelet transform gives a measure of correlation between the signal and the wavelet. A larger value of wavelet coefficient means a larger correlation between the signal and wavelet. Therefore by looking at the wavelet coefficients that are produced when an ERP signal is analyzed by different wavelet candidates, it is possible to determine the best wavelet basis for a particular ERP component detection application.

Given the shape and spectral profile of the N170 ERP, wavelets such as Daubechies 6, Daubechies 7, Daubechies 8, Symlet 5, Biorthogonal 3.9  are possible candidates. The CWT coefficients of these wavelets at specific scales were examined with respect to the representative averaged N170 ERP shown in Figure 3f. 




The large value of CWT coefficients revealed that Symlet 5 had the best match with the averaged N170 ERP.

\begin{figure}[!h]
\includegraphics{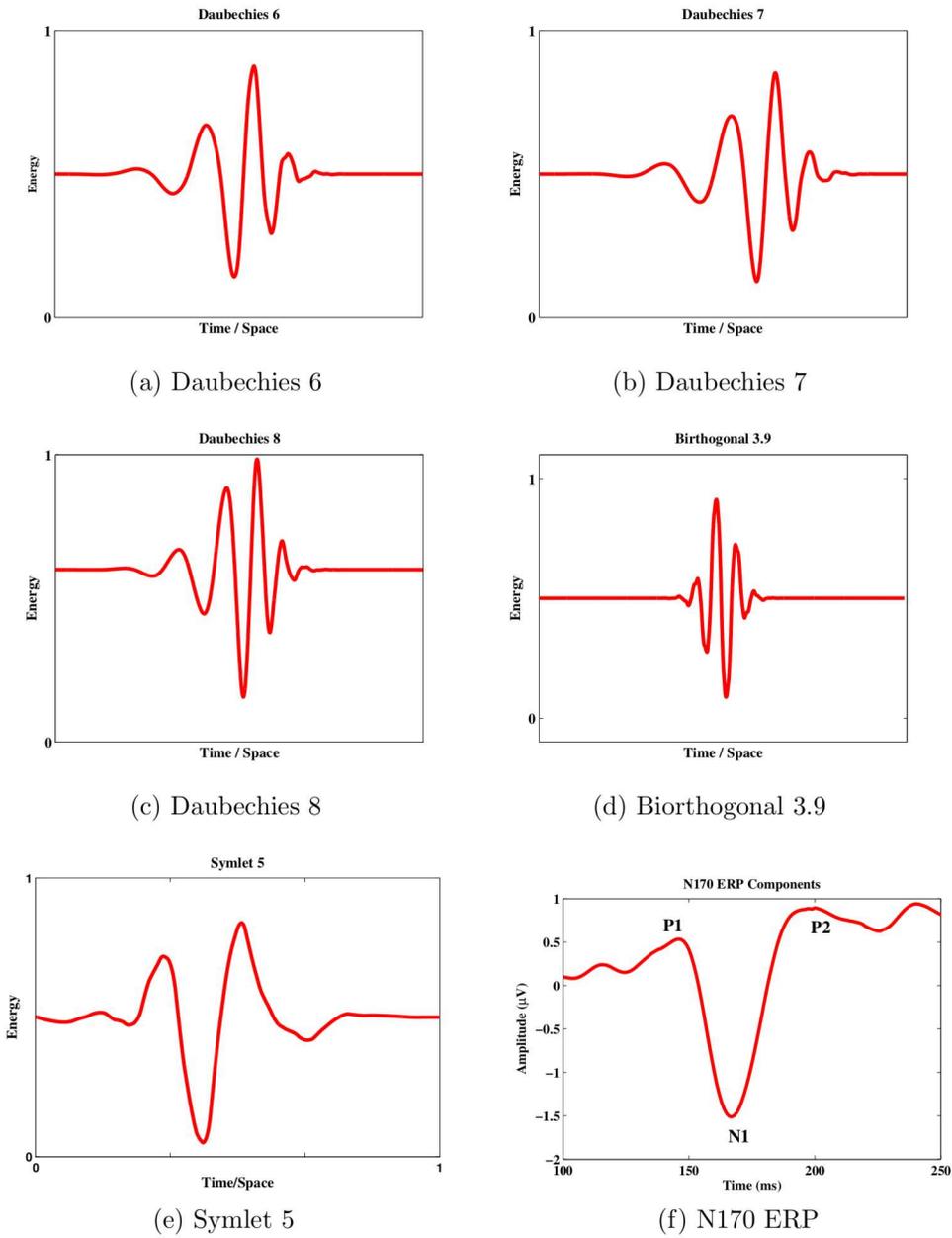}
\caption{The Symlet 5 wavelet has the greatest similarity in terms of spectral content to that of the N170 ERP complex. This is also evident from their morphological similarity shown here.}\label{fig3}
\end{figure}

The N170 ERP component has characteristic component peaks, namely P1, N1 and P2. The Symlet 5 has similar features in its profile which is why it matched the component better than others. This is illustrated in Figure \ref{fig3}.  


In the present study, to determine the best scale, the occipito-temporal signals from all 10 subjects who had prominent P1, N1 and P2 peaks comprising the N170 component were segmented manually. The segmented data were averaged for each channel across all the subjects to derive an averaged ERP. A full CWT was then done on each of the averaged segments. The scale at which the wavelet coefficients gives a maximum value (i.e., at which spectral matching is a maximum) was identified as the best scale for detecting N170 ERPs.

\begin{figure}[!h]
\centering
\includegraphics{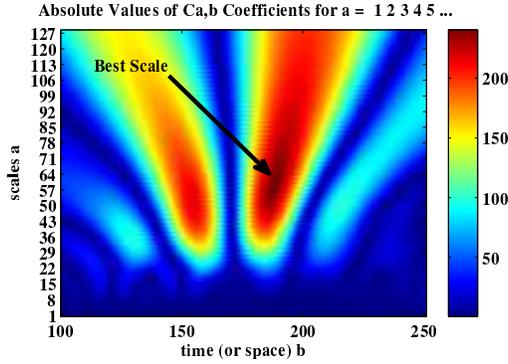}
\caption{The CWT gives large coefficient values at regions where the scaling of the wavelet matches the structure in the input signal. The corresponding scale and time are marked by the darkest region in the time-scale Full CWT plot shown above.}\label{fig4}
\end{figure}

The sum of absolute values of the coefficients at each scale for every channel is then computed as:
\begin{equation}
S_{n}(a) = \sum abs[C_{n}(a)] \label{eq5}
\end{equation}

Finally, the scale at which the value of the sum $\mathit{S_{n}(a)}$ is maximum  is chosen as the best match \citep{rafee} for the specific channel $\mathit{n}$. 
Due to the characteristic nature of the averaged ERP, we observed that the values for best scale varied across the channels. On  average it was noted that a scale of 65 was the best scale returned for the different channels. Figure \ref{fig4} shows the range of matching scales for the N170 ERP in this study.

\subsection{Detection Algorithm}
\begin{figure}[!h]
\begin{center}
\includegraphics{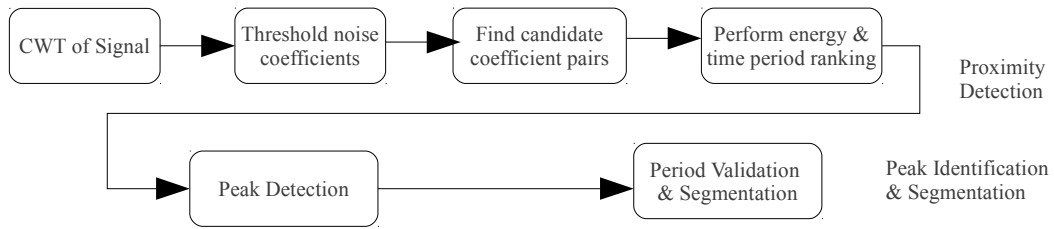}  
\caption{Detection Algorithm}\label{fig5}
\end{center}
\end{figure}

The detection and isolation of N170 is carried out in two stages, namely, Proximity and Peak detection tasks as shown in Figure \ref{fig5}. 

The algorithm begins with the CWT of the input signal to detect the candidate locations for the N170 ERP and then marks the exact location of the ERP after a validation procedure. In a fully automated algorithm based on the CWT, the best scale of analysis should also be chosen automatically. To resolve this issue we added an additional decision function prior to the main algorithm to determine the scale of analysis. 

\begin{figure}[!h]
\includegraphics{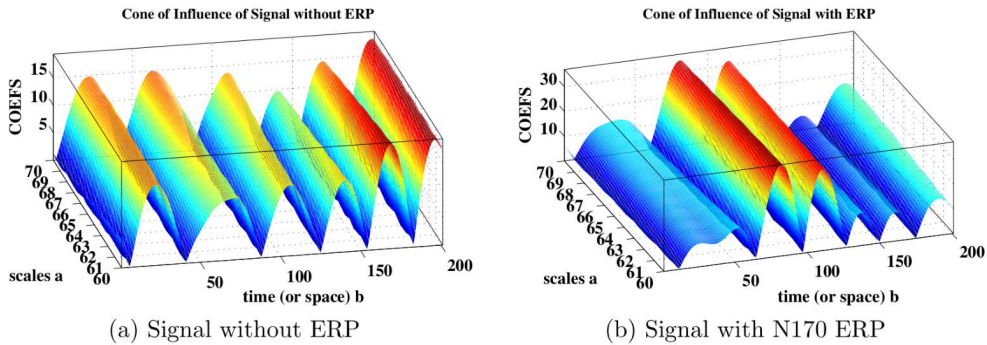}
\caption{(a) shows the spread of higher coefficients over a large area for different scales for a single trial EEG signal with just noise and no ERP. (b) shows the localization of higher coefficients over different scales for a signal with an N170 ERP.}\label{fig6}
\end{figure}

The decision function now tests for the presence of large wavelet coefficient localization across the specified range of scales. This region is called the cone of influence \citep{terrence}. In most of the cases where there is a highlighted anomaly similar to the ERP in the signal, the cone of influence  will be narrow and well defined. See Figure \ref{fig6}.

Once the cone of influence is identified, the scale corresponding to the largest wavelet coefficient is chosen as the best scale of analysis. Selection of the best scale becomes a challenging issue when dealing with single trial EEG data. The original ERP will be modulated to a certain amount by the high alpha components and other noise sources. 


\subsubsection{Proximity Detection} 
The proximity detection stage finds a candidate location for an N170 component based on the wavelet coefficients.
The signal is first preprocessed using a low pass  filter designed with cut off frequency 65 Hz and stop band attenuation of 25 Hz to remove high frequency spikes within the data. These parameters are decided from the power spectrum of the signal.

After the preprocessing stage, CWT is done on the signal's $\mathit{n^{th}} $ channel with the Symlet 5 wavelet at  the specified scale $a$ to generate wavelet coefficients $\mathit{C_{n}(a)}$. A threshold of wavelet coefficient $\mathit{C_\tau}$ is set such that all coefficients that satisfy the condition
\begin{equation}
\mathrm{C_{n}(a)} \geqslant  C_\tau \label{eq6}
\end{equation}
are retained in the process. The value of $\mathit{C_\tau}$ is set to 50\% of the average wavelet coefficient corresponding to the N170 peak of all the trials used to determine the range of scales in Section \ref{scalesel}.
Suppressing all remaining coefficients reduces the chance for false detection in the steps that follow. If no coefficient with a value greater than $\mathit{C_\tau}$ is found in the data, the algorithm reports the absence of an ERP and exits without going to the remaining processing stages. 

The maxima of positive and negative coefficients appear in pairs with the negative coefficient preceding the positive coefficient where the N170 complex is present.  When the wavelet is convolved with the signal on the time axis, there will be a point of minimum correlation and maximum correlation. The wavelet function being slightly asymmetric \citep{daube}, the points of correlation will be shifted with respect to the alignment of the largest peak in the N170 versus the wavelet by a certain factor. This is the shift $\delta_{k}$ in terms of sample number introduced by the wavelet. Considering the band limited nature of the wavelet and the ERP, the shift will stay within a specific range.

\begin{figure}[!h]
\centering
\includegraphics{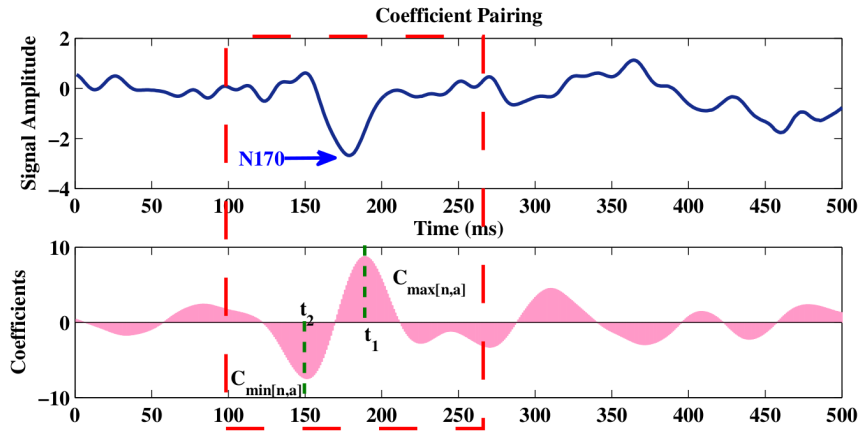} 
\caption{At regions where the N170 ERP complex is detected, the wavelet transform gives a pair of negative and positive coefficients. The strength of these coefficients depends on how well the match is.}\label{fig7}
\end{figure}

According to the basic theory of wavelets, larger energy distribution corresponds to areas where an anomaly is present. In an ideal case this should be where the N170 ERP occurs. But in actual practice, due to noise and other factors, there could be multiple segments similar to N170 in an EEG. If the algorithm finds more than two coefficients satisfying the condition in equation \ref{eq6}, all candidate coefficient pairs for which the negative coefficient precedes the positive coefficient are identified and ranked on the basis of the energy of the coefficients. 
To differentiate the N170 component from seemingly similar segments, we add a constraint that the time period between the maximum positive coefficient and negative coefficient must be such that,
\begin{equation}
 \mathrm{t_1 - t_2} \approx T_{N170} \label{eq7}
\end{equation}
where $\mathit{T_{N170}}$ is the approximate time period observed for N170 coefficient pairs in the data. See Figure \ref{fig7}. In our data, this time period is around 60ms - 88ms for single-trial EEG data. This is the asymmetry based measure derived from the wavelet coefficient.

\begin{figure}[!h]
\centering
\includegraphics{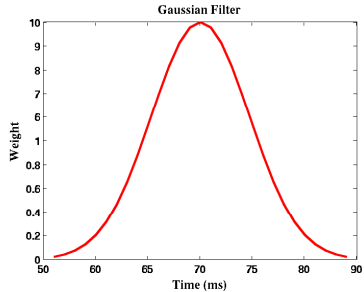}  
\caption{In most subjects, the positive and negative coefficients corresponding to the ERP complex are separated by about 70ms. However, this is not a magic number and may vary slightly across subjects and observations. We thus used a Gaussian Weight function such that any detection that does not match in time may be discarded. This reduces false detection resulting from noise when SNR is low.} \label{fig8}
\end{figure}

We use a Gaussian weight function centered around 70 ms to give relative weighting to each of the detected segments as shown in Figure \ref{fig8}. Thus if the coefficient that is a maximum has a $\mathit{T_{N170}}$ of about 70ms, it gets the highest ranking in the detection procedure. 

\begin{figure}[!h]
\centering
\includegraphics{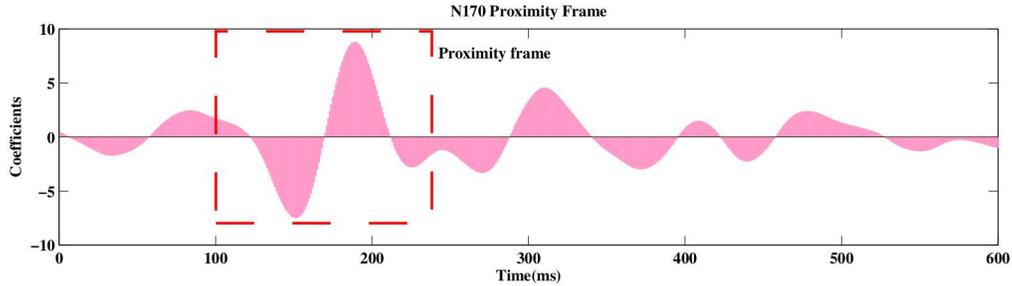} 
\caption{We use a Proximity Detection algorithm to pick up the best coefficients after applying the Gaussian weight function. The boundaries of the proximity frame  spans the maximum value of $\mathit{T_{N170}}$. This procedure discards fake detections that do not match in either spectral content or scale, thereby improving the detection accuracy of the algorithm.}\label{fig9}
\end{figure}

In other words, the best coefficient pair will have maximum energy as well as the best value of $\mathit{T_{N170}}$. This strategy will reduce the chances of picking up false candidates having maximum energy distribution. Figure \ref{fig9} shows an example of a selected coefficient pair within the proximity frame. 

The time index values of successful detection candidates are saved and passed on to the next stage of the algorithm.


\subsubsection{Peak Identification and Segmentation}

The second stage of the algorithm constitutes a peak detection mechanism along with a final segmentation and validation. The peak identification algorithm basically detects the P1, N1 and P2 components of the N170 complex.
A forward and backward slope detection mechanism is the heart of the peak identification algorithm. The peak detection first picks the N1 component which is the main component of N170. Once N1 is identified, the minor components P1 and P2 are detected. To minimize the effect of picking local peaks due to noise, a minimum peak period validation is performed. In this validation step, the time period between the candidate P1, N1 and P2 are checked. A detection is recorded if the time period between the 3 complexes satisfy the time period of the standard N170 positive and negative peaks.
Once the peaks are identified, a final check on the period of the ERP is done to verify that it meets the N170 total period criterion. If it fails, the peak identification is re-run with modified parameters to recalculate the peaks. The algorithm finally marks the start and end positions of the segmented N170 ERP. 

This final stage of the algorithm also makes sure that the basic morphology of the N170 component which are P1, N1 and P2 are conserved while segmenting. The segment is padded with a short period of 10ms at the start and end of the extracted data. 


\section{Results} \label{results}
We tested the algorithm on the occipito-posterior sites of the single trial data set of 10 subjects. Along with the EEG data, the wavelet name and sampling rate were given as input arguments to the algorithm. The test was done in two stages,  namely	1) verifying	 that the algorithm does not detect false N170 components	 in resting EEG, and 2) 	comparing the ability of the algorithm to detect genuine single-trial N170 components associated with stimulus 
presentation against two commonly used wavelet detection algorithms.

\subsection{Resting EEG data}

\begin{figure}[!h]
\centering
\includegraphics[scale =.45]{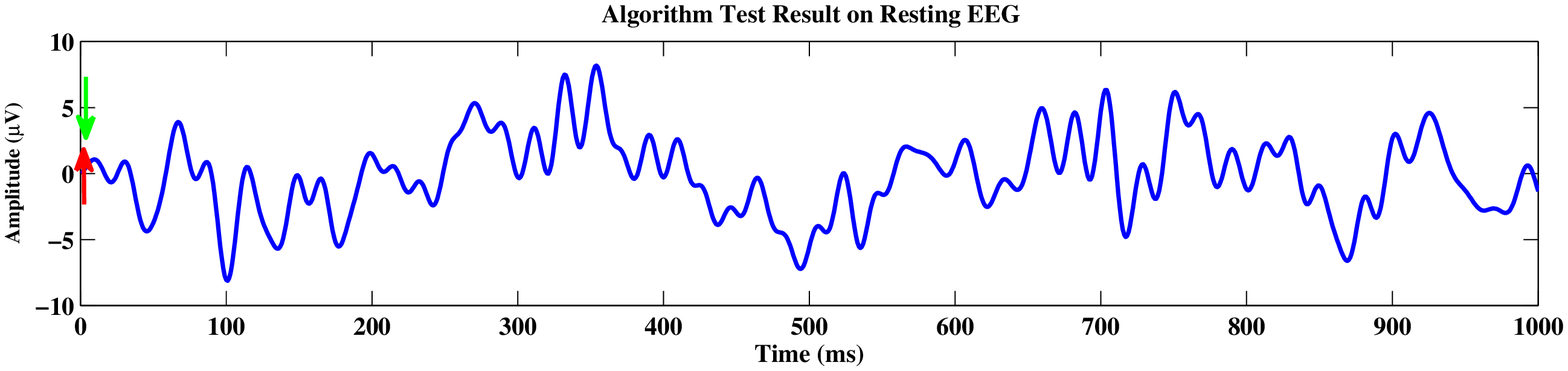}
\caption{An example used to demonstrate that erroneous detections are not picked up by the algorithm is shown. Green and red arrows are used to mark the start and end of the N170 component. When no ERP is found, this is reported by the arrows positioned one on top of the other as shown.}\label{fig10}
\end{figure}

It is desired that the ERP detection algorithm identify only real ERPs and not patterns that look similar to an ERP. To verify this, our algorithm was tested on 400 samples of resting EEG data without N170 ERP components. When a detection is made, the algorithm marks the start and end of the ERP with green and red arrow marks. When the algorithm is not able to find an ERP in the data, the arrows are placed at the beginning of the data stream, pointing at each other. The algorithm did not identify an N170 component in any of the resting EEG segments. A sample data segment used for the study is shown in Figure \ref{fig10}.

\subsection{Single trial EEG}
\begin{figure}[!h]
\centering
\includegraphics{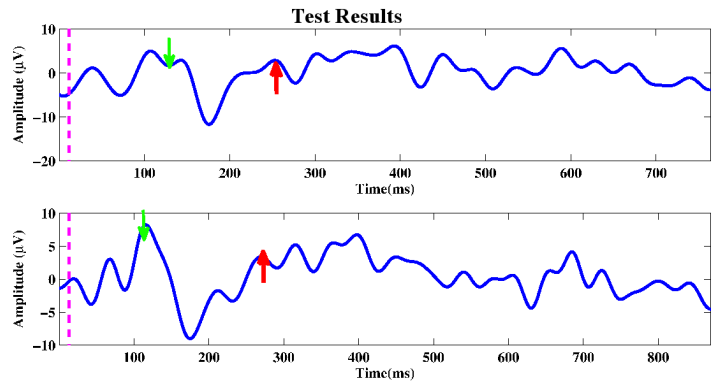} 
\caption{Two examples of N170 ERPs in single trial data that  were correctly detected by our algorithm are shown. The dashed line shows the onset of the stimulus. The two examples are data from the O2 channel of two different subjects for which the best wavelet scales were different for the two subjects. }
\label{fig11}
\end{figure}

The algorithm was tested for its ability to detect the N170 component in the single-trial EEG data that had been preprocessed to remove artifacts caused by eye movements, ECG spikes and muscle activity. For the single trial test, Twelve hundred single-trial ERP segments from the occipito-temporal channels of all the 10 subjects were used. Out of the 1200 trials, 600 were ERPs to face stimuli and the other 600 were ERPs to non-face stimuli (including house and checkerboard stimuli). The algorithm detected the presence of single-trial N170 ERPs to faces in the 600 positive samples with an accuracy of 96.7\% and the accuracy with which it detected the absence of face-evoked N170 ERPs  was 86\%. With the total 1200 samples the overall accuracy was 91.33\%. Figure \ref{fig11} shows two examples in which the N170 ERPs were correctly detected by our algorithm. 

To evaluate how well our algorithm performs, it was compared against the spikelet technique proposed by \cite{rodri} and t-CWT proposed by \cite{bostanov2}. The Spikelet constructs wavelets matching a specified signal by deriving a system of linear equations from the Daubechies transform filter coefficients. A matching wavelet with the spikelet algorithm was constructed to match the N170 ERP and was used for detection. The t-CWT method uses a  combination of t-statistics and wavelet scalogram called the t-value scalogram to extract features of a particular ERP. This measure can be used for ERP detection. The t-CWT method was modified such that the wavelet filter coefficients of the Symlet system were used to calculate the scalogram to detect the ERP. The results are tabulated in Table \ref{alg}. 

The first column in the table shows the three methods for ERP detection. The third column shows the number of predictions of the corresponding ERP type in column 2. The detection accuracy of each type of ERP is given in the fourth column and the last column gives the overall accuracy of each algorithm to detect the presence/absence of face-evoked ERP. As Table \ref{alg} shows, the proposed asymmetry based detection algorithm is more accurate in identifying ERP structures in the data. 

\begin{center}
\begin{table}[!htbp]
\begin{tabular}{|l|c||cc||c|c|}
\hline 
Method& Real & Predicted & Predicted as &Detection&Overall \\ 
 & &as N170 & non-N170 &Accuracy& Accuracy\\ \hline \hline
Matching & N170 & \multicolumn{1}{c|}{190} & \multicolumn{1}{c||}{410} & 31.7\% &\\ \hline
& Non-N170 & \multicolumn{1}{c|}{278} & \multicolumn{1}{c||}{322}&53.7\% & 42.66 \%  \\ \hline\hline
t-CWT& N170 & \multicolumn{1}{c|}{512} & \multicolumn{1}{c||}{88} & 85.3\%&\\ \hline
& Non-N170 & \multicolumn{1}{c|}{112} & \multicolumn{1}{c||}{488}& 81.3\% &83.3 \%  \\ \hline \hline
Asymmetry& N170 & \multicolumn{1}{c|}{580} & \multicolumn{1}{c||}{20} & 96.7\% &\\ \hline
& Non-N170 & \multicolumn{1}{c|}{84} & \multicolumn{1}{c||}{516} &86.0\% & 91.33\% \\ \hline
\end{tabular}
\caption{Comparison of the  ERP detection accuracy of the asymmetry method  with two other algorithms using 1200 instances of single trial EEG data. }
\label{alg}
\end{table} 
\end{center}

\begin{figure}[!h]
\centering
\includegraphics{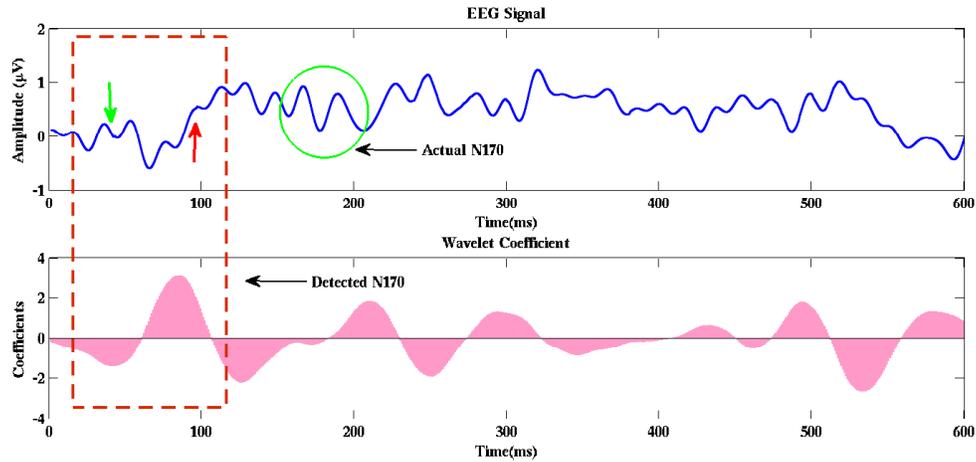}
\caption{A typical failed case is shown. This failure resulted due to the distortion caused by the noise in the spectral components of the ERP complex and the resulting reduction in the strength of the coefficient of the actual ERP.}\label{fig12}
\end{figure}

Detection failures were separately investigated in detail. Most of them occurred in cases where the  original ERPs are severely distorted by high alpha components from noisy patterns that had spectral components similar to N170. Figure \ref{fig12} illustrates how a fake pattern may generate larger wavelet coefficients above the detection threshold limit to result in the false detection of the ERP.

The time period check could not save the situation because the fake candidate pair  matched  the expected time period and gave a larger value for the wavelet coefficient.  

 
With N170 single trial data, the majority of the failed detection cases were similar to the example in Figure \ref{fig12} where the prominent features of the actual ERP were distorted by noise in the signal. This distortion reduced the value of wavelet coefficient and the time period of the actual N170 coefficient pair did not fall within the set of optimum values. The noise fluctuation had the value of wavelet coefficient and time period of an actual N170 ERP. In all those cases, the other detection algorithms also failed to correctly identify the ERP component. 


\section{Discussion}
This study demonstrates that the wavelet coefficients at different scales and the corresponding phase shift are unique for a particular ERP component, and can be used for detection of event-related potentials in single trial data.

The efficiency of the algorithm was tested by detecting N170 ERPs in single trial data. The algorithm robustly detected the N170 ERPs except in the few situations where  the noise component presented a more faithful match to the actual ERP component or distorted the prominent features of the actual ERP component. 

The detection of wide variety of single-trial ERP components has many potential applications, including to reveal fluctuations in the information content across trials\citep{rouss}, to disentangle the contribution of single-trial amplitudes and latencies to changes in mean ERPs\citep{navjas} and to model covariation across recording modalities, for instance EEG and fMRI\citep{nguyen}.

Nevertheless, there are limitations to the types of ERP components that the algorithm can detect with high accuracy.
The algorithm may not be useful for detecting slow wave ERP components. This is because the asymmetry measure requires that the selected ERP component should be oscillatory. When wavelets with asymmetric response are convolved with slow wave ERPs, the induced  group delay will be very minimal making it impossible to obtain the asymmetry measure. Therefore the asymmetry measure may fail due to the lack of a good wavelet match.

We have shown that the algorithm gives good results in offline mode. The offline detection of a specific ERP component on single trials is an extremely useful procedure with many applications, both in basic neuroscience research comparing neural responses to cognitive information processing in one condition versus another, and in clinical neuroscience research designed to determine symptoms associated with compromised brain functions. In both cases, standard ERP methods that depend on averaging many single trials in order to eliminate EEG responses not associated with the stimulus manipulation also end up eliminating important information only available when single trials are scored, namely, the degree to which the brain response is consistent across trials. For example, intra-individual variability of behavioral responses \citep{mcdonald} and of ERPs \citep{sego} relate to important cognitive and neural control functions, and potentially inform us of the progress of variable disease processes \citep{kiiski}. However, the methods currently available to detect this variability are relatively crude. One such method is to apply a severe low-pass filter that eliminates everything except a low-frequency large component such as the P300\citep{sego}.Others have specific parameter requirements, such as independent component analysis  (ICA) which requires having an adequate number of electrode channels (e.g., 64 sites) and having stationery signals \citep{kiiski}. The algorithm presented here permits the scoring on single-trials of the amplitude and latency (with respect to a stimulus-onset time) of an ERP component, thus yielding a latency-consistency measure. It also permits one to examine the statistical reliability of an amplitude difference (of a specific ERP component) between two conditions for each research participant, thus removing the need to assimilate to the error term differences in EEG response magnitude across individuals. This is a useful aspect of the algorithm that has great potential for cognitive and clinical neuroscience. 

The implementation of this algorithm in real time potentially offers opportunities to improve brain computer interfaces for augmenting human cognitive and motor abilities and for advancing the field of neuroprosthetics. This is a topic for future research. There are a number of limitations that must be overcome in order to achieve a real time implementation. First, the algorithm is based on the CWT which in general is a computationally expensive technique, and because the algorithm must perform multiple CWTs during a single detection, processor speed may be a limiting factor to real-time implementation. Currently the algorithm execution takes nearly 80ms on a PC with 2.2GHz Intel i3 processor and 2GB RAM. Second, artifacts caused by eye movement and muscle movement were removed from the data used in this study by an independent components analysis (ICA), a computationally intensive off-line procedure. However real time EEG data are highly contaminated with such extracerebral artifacts and the detection performance of the algorithm under conditions of currently available real-time artifact rejection remains to be determined in future research.

\section{Conclusion}
We have introduced a novel wavelet-based algorithm based on wavelet asymmetry for detecting single-trial oscilatory ERP components and have shown that it provides more accurate off-line detection than two common previous methods. The algorithm is accurate even when moderate noise is present in the data.

\section*{Acknowledgments}
The authors wish to  thank Jill Munro for developing the ERP stimuli used in this paper and Mrs. Elizabeth John for proofreading the document. The authors wish to thank the anonymous reviewers for their valuable and detailed
comments that greatly improved the quality of the manuscript.

\bibliographystyle{elsarticle-harv} \biboptions{authoryear}

\begin{thebibliography}{42}
\expandafter\ifx\csname natexlab\endcsname\relax\def\natexlab#1{#1}\fi
\providecommand{\url}[1]{\texttt{#1}}
\providecommand{\href}[2]{#2}
\providecommand{\path}[1]{#1}
\providecommand{\DOIprefix}{doi:}
\providecommand{\ArXivprefix}{arXiv:}
\providecommand{\URLprefix}{URL: }
\providecommand{\Pubmedprefix}{pmid:}
\providecommand{\doi}[1]{\href{http://dx.doi.org/#1}{\path{#1}}}
\providecommand{\Pubmed}[1]{\href{pmid:#1}{\path{#1}}}
\providecommand{\bibinfo}[2]{#2}
\ifx\xfnm\relax \def\xfnm[#1]{\unskip,\space#1}\fi
\bibitem[{Abootalebi et~al.(2006)Abootalebi, Moradi and Khalilzadeh}]{vahid}
\bibinfo{author}{Abootalebi, V.}, \bibinfo{author}{Moradi, M.H.},
  \bibinfo{author}{Khalilzadeh, M.A.}, \bibinfo{year}{2006}.
\newblock \bibinfo{title}{{A comparison of methods for ERP assessment in a
  P300-based GKT}}.
\newblock \bibinfo{journal}{International Journal of Psychophysiology}
  \bibinfo{volume}{62}, \bibinfo{pages}{309--320}.
\bibitem[{Barrett(2000)}]{barret}
\bibinfo{author}{Barrett, G.}, \bibinfo{year}{2000}.
\newblock \bibinfo{title}{{Clinical application of event-related potentials in
  dementing illness: issues and problems}}.
\newblock \bibinfo{journal}{International Journal of psychophysiology}
  \bibinfo{volume}{37}, \bibinfo{pages}{49--53}.
\bibitem[{Bentin et~al.(1996)Bentin, Allison, Puce, Perez and
  McCarthy}]{bentin}
\bibinfo{author}{Bentin, S.}, \bibinfo{author}{Allison, T.},
  \bibinfo{author}{Puce, A.}, \bibinfo{author}{Perez, E.},
  \bibinfo{author}{McCarthy, G.}, \bibinfo{year}{1996}.
\newblock \bibinfo{title}{{Electrophysiological studies of face perception in
  humans}}.
\newblock \bibinfo{journal}{Journal of Cognitive Neuroscience}
  \bibinfo{volume}{8}, \bibinfo{pages}{551--565}.
\bibitem[{Birbaumer et~al.(2008)Birbaumer, Murguialday and Cohen}]{birbaumer}
\bibinfo{author}{Birbaumer, N.}, \bibinfo{author}{Murguialday, A.R.},
  \bibinfo{author}{Cohen, L.}, \bibinfo{year}{2008}.
\newblock \bibinfo{title}{{Brain-computer interface in paralysis.}}
\newblock \bibinfo{journal}{Current Opinion in Neurology} \bibinfo{volume}{21},
  \bibinfo{pages}{634--638}.
\bibitem[{Blanco et~al.(1998)Blanco, Figliola, Quiroga, Rosso and
  Serrano}]{blanco}
\bibinfo{author}{Blanco, S.}, \bibinfo{author}{Figliola, A.},
  \bibinfo{author}{Quiroga, R.Q.}, \bibinfo{author}{Rosso, O.A.},
  \bibinfo{author}{Serrano, E.}, \bibinfo{year}{1998}.
\newblock \bibinfo{title}{{Time-frequency analysis of electroencephalogram
  series. III. Wavelet packets and information cost function}}.
\newblock \bibinfo{journal}{Physical Review E} \bibinfo{volume}{57},
  \bibinfo{pages}{932}.
\bibitem[{Bostanov(2004)}]{bostanov2}
\bibinfo{author}{Bostanov, V.}, \bibinfo{year}{2004}.
\newblock \bibinfo{title}{{BCI competition 2003-data sets Ib and IIb: feature
  extraction from event-related brain potentials with the continuous wavelet
  transform and the t-value scalogram}}.
\newblock \bibinfo{journal}{IEEE Transactions Biomedical Engineering}
  \bibinfo{volume}{51}, \bibinfo{pages}{1057--1061}.
\bibitem[{Bostanov and Kotchoubey(2004)}]{bostanov1}
\bibinfo{author}{Bostanov, V.}, \bibinfo{author}{Kotchoubey, B.},
  \bibinfo{year}{2004}.
\newblock \bibinfo{title}{{Recognition of affective prosody: Continuous wavelet
  measures of event-related brain potentials to emotional exclamations}}.
\newblock \bibinfo{journal}{Psychophysiology} \bibinfo{volume}{41},
  \bibinfo{pages}{259--268}.
\bibitem[{Brechet et~al.(2007)Brechet, Lucas, Doncarli and Farina}]{brechet}
\bibinfo{author}{Brechet, L.}, \bibinfo{author}{Lucas, M.F.},
  \bibinfo{author}{Doncarli, C.}, \bibinfo{author}{Farina, D.},
  \bibinfo{year}{2007}.
\newblock \bibinfo{title}{{Compression of biomedical signals with mother
  wavelet optimization and best-basis wavelet packet selection}}.
\newblock \bibinfo{journal}{IEEE Transactions on Biomedical Engineering}
  \bibinfo{volume}{54}, \bibinfo{pages}{2186--2192}.
\bibitem[{Browne and Cutmore(2002)}]{browne}
\bibinfo{author}{Browne, M.}, \bibinfo{author}{Cutmore, T.R.H.},
  \bibinfo{year}{2002}.
\newblock \bibinfo{title}{{Low-probability event-detection and separation via
  statistical wavelet thresholding: an application to psychophysiological
  denoising}}.
\newblock \bibinfo{journal}{Clinical Neurophysiology} \bibinfo{volume}{113},
  \bibinfo{pages}{1403--1411}.
\bibitem[{Chapa(1995)}]{chapathesis}
\bibinfo{author}{Chapa, J.O.}, \bibinfo{year}{1995}.
\newblock \bibinfo{title}{Matched Wavelet Construction and Its Application to
  Target Detection}.
\newblock Ph.D. thesis. Center for Imaging, Rochester Institute of Technology.
\bibitem[{Chapa and Rao(2000)}]{chapa}
\bibinfo{author}{Chapa, J.O.}, \bibinfo{author}{Rao, R.M.},
  \bibinfo{year}{2000}.
\newblock \bibinfo{title}{{Algorithms for designing wavelets to match a
  specified signal}}.
\newblock \bibinfo{journal}{IEEE Transactions on Signal Processing}
  \bibinfo{volume}{48}, \bibinfo{pages}{3395--3406}.
\bibitem[{Ciniburk and Mautner(2008)}]{jind}
\bibinfo{author}{Ciniburk, J.}, \bibinfo{author}{Mautner, P.},
  \bibinfo{year}{2008}.
\newblock \bibinfo{title}{{Suitability of Huang Hilbert Transformation for ERP
  detection}}, in: \bibinfo{booktitle}{Proceedings of 9th International PhD
  Workshop on Systems and Control: Young Generation Viewpoint}.
\bibitem[{Daubechies(1992)}]{daube}
\bibinfo{author}{Daubechies, I.}, \bibinfo{year}{1992}.
\newblock \bibinfo{title}{{Ten lectures on wavelets}}.
\newblock \bibinfo{publisher}{Society for industrial and applied mathematics}.
\bibitem[{Demiralp et~al.(1999)Demiralp, Ademoglu, Sch\"{u}rmann, Basar-Eroglu
  and Basar}]{demir}
\bibinfo{author}{Demiralp, T.}, \bibinfo{author}{Ademoglu, A.},
  \bibinfo{author}{Sch\"{u}rmann, M.}, \bibinfo{author}{Basar-Eroglu, C.},
  \bibinfo{author}{Basar, E.}, \bibinfo{year}{1999}.
\newblock \bibinfo{title}{{Detection of P300 waves in single trials by the
  wavelet transform (WT)}}.
\newblock \bibinfo{journal}{Brain and Language} \bibinfo{volume}{66},
  \bibinfo{pages}{108--128}.
\bibitem[{Desjardins and Segalowitz(2013)}]{james}
\bibinfo{author}{Desjardins, J.A.}, \bibinfo{author}{Segalowitz, S.J.},
  \bibinfo{year}{2013}.
\newblock \bibinfo{title}{{Deconstructing the early visual electrocortical
  responses to face and house stimuli}}.
\newblock \bibinfo{journal}{Journal of Vision} \bibinfo{volume}{13}.
\bibitem[{Fatourechi et~al.(2004)Fatourechi, Mason, Birch and Ward}]{fatour}
\bibinfo{author}{Fatourechi, M.}, \bibinfo{author}{Mason, S.G.},
  \bibinfo{author}{Birch, G.E.}, \bibinfo{author}{Ward, R.K.},
  \bibinfo{year}{2004}.
\newblock \bibinfo{title}{{A wavelet-based approach for the extraction of event
  related potentials from EEG}}, in: \bibinfo{booktitle}{Proceeding of IEEE
  International Conference on Acoustic Speech Signal Processing (ICASSP'04)},
  \bibinfo{organization}{IEEE}. pp. \bibinfo{pages}{ii----737}.
\bibitem[{Flanders(2002)}]{flanders}
\bibinfo{author}{Flanders, M.}, \bibinfo{year}{2002}.
\newblock \bibinfo{title}{{Choosing a wavelet for single-trial EMG}}.
\newblock \bibinfo{journal}{Journal of Neuroscience Methods}
  \bibinfo{volume}{116}, \bibinfo{pages}{165--177}.
\bibitem[{Ford et~al.(1994)Ford, White, Lim and Pfefferbaum}]{ford}
\bibinfo{author}{Ford, J.M.}, \bibinfo{author}{White, P.},
  \bibinfo{author}{Lim, K.O.}, \bibinfo{author}{Pfefferbaum, A.},
  \bibinfo{year}{1994}.
\newblock \bibinfo{title}{{Schizophrenics have fewer and smaller P300s: a
  single-trial analysis}}.
\newblock \bibinfo{journal}{Biological Psychiatry} \bibinfo{volume}{35},
  \bibinfo{pages}{96--103}.
\bibitem[{Graimann et~al.(2004)Graimann, Huggins, Levine and
  Pfurtscheller}]{bern}
\bibinfo{author}{Graimann, B.}, \bibinfo{author}{Huggins, J.E.},
  \bibinfo{author}{Levine, S.P.}, \bibinfo{author}{Pfurtscheller, G.},
  \bibinfo{year}{2004}.
\newblock \bibinfo{title}{{Toward a direct brain interface based on human
  subdural recordings and wavelet-packet analysis}}.
\newblock \bibinfo{journal}{IEEE Transactions on Biomedical Engineering}
  \bibinfo{volume}{51}, \bibinfo{pages}{954--962}.
\bibitem[{Guido et~al.(2006)Guido, Slaets, K\"{o}berle, Almeida and
  Pereira}]{rodri}
\bibinfo{author}{Guido, R.C.}, \bibinfo{author}{Slaets, J.F.W.},
  \bibinfo{author}{K\"{o}berle, R.}, \bibinfo{author}{Almeida, L.O.B.},
  \bibinfo{author}{Pereira, J.C.}, \bibinfo{year}{2006}.
\newblock \bibinfo{title}{{A new technique to construct a wavelet transform
  matching a specified signal with applications to digital, real time, spike,
  and overlap pattern recognition}}.
\newblock \bibinfo{journal}{Digital Signal Processing} \bibinfo{volume}{16},
  \bibinfo{pages}{24--44}.
\bibitem[{Herrera et~al.(2000)Herrera, Sun, Charles, Dahl, Ryan and
  Sclabassi}]{herrera}
\bibinfo{author}{Herrera, R.E.}, \bibinfo{author}{Sun, M.},
  \bibinfo{author}{Charles, P.J.}, \bibinfo{author}{Dahl, R.E.},
  \bibinfo{author}{Ryan, N.D.}, \bibinfo{author}{Sclabassi, R.J.},
  \bibinfo{year}{2000}.
\newblock \bibinfo{title}{{Removal of non-white noise from single trial
  event-related EEG signals using soft-thresholding}}, in:
  \bibinfo{booktitle}{Proceedings of the 22nd Annual International Conference
  of the IEEE, Engineering in Medicine and Biology Society},
  \bibinfo{organization}{IEEE}. pp. \bibinfo{pages}{793--795}.
\bibitem[{Huang and Shen(2005)}]{hhtr}
\bibinfo{author}{Huang, N.E.}, \bibinfo{author}{Shen, S.S.},
  \bibinfo{year}{2005}.
\newblock \bibinfo{title}{{Hilbert-Huang transform and its applications}}.
\newblock \bibinfo{publisher}{World Scientific}.
\bibitem[{Jiang et~al.(2007)Jiang, Chao, Chiu, Lee, Tseng and Lin}]{joe}
\bibinfo{author}{Jiang, J.A.}, \bibinfo{author}{Chao, C.F.},
  \bibinfo{author}{Chiu, M.J.}, \bibinfo{author}{Lee, R.G.},
  \bibinfo{author}{Tseng, C.L.}, \bibinfo{author}{Lin, R.},
  \bibinfo{year}{2007}.
\newblock \bibinfo{title}{An automatic analysis method for detecting and
  eliminating ecg artifacts in eeg}.
\newblock \bibinfo{journal}{Computers in Biology and Medicine}
  \bibinfo{volume}{37}, \bibinfo{pages}{1660--1671}.
\bibitem[{Kalayci et~al.(1994)Kalayci, Ozdamar and Erdol}]{kala}
\bibinfo{author}{Kalayci, T.}, \bibinfo{author}{Ozdamar, O.},
  \bibinfo{author}{Erdol, N.}, \bibinfo{year}{1994}.
\newblock \bibinfo{title}{{The use of wavelet transform as a preprocessor for
  the neural network detection of EEG spikes}}, in:
  \bibinfo{booktitle}{Proceedings of the 1994 IEEE conference Southeastcon'94
  Creative Technology Transfer - A Global Affair},
  \bibinfo{organization}{IEEE}. pp. \bibinfo{pages}{1--3}.
\bibitem[{Kiiski et~al.(2012)Kiiski, Reilly, Lonergan, Kelly, O'Brien,
  Kinsella, Bramham, Burke, Donnchadha, Nolan and Others}]{kiiski}
\bibinfo{author}{Kiiski, H.}, \bibinfo{author}{Reilly, R.B.},
  \bibinfo{author}{Lonergan, R.}, \bibinfo{author}{Kelly, S.},
  \bibinfo{author}{O'Brien, M.C.}, \bibinfo{author}{Kinsella, K.},
  \bibinfo{author}{Bramham, J.}, \bibinfo{author}{Burke, T.},
  \bibinfo{author}{Donnchadha, S.O.}, \bibinfo{author}{Nolan, H.},
  \bibinfo{author}{Others}, \bibinfo{year}{2012}.
\newblock \bibinfo{title}{{Only Low Frequency Event-Related EEG Activity Is
  Compromised in Multiple Sclerosis: Insights from an Independent Component
  Clustering Analysis}}.
\newblock \bibinfo{journal}{PLoS One} \bibinfo{volume}{7},
  \bibinfo{pages}{e45536}.
\bibitem[{Lim et~al.(1995)Lim, Akay and Daubenspek}]{lim}
\bibinfo{author}{Lim, L.M.}, \bibinfo{author}{Akay, M.},
  \bibinfo{author}{Daubenspek, J.A.}, \bibinfo{year}{1995}.
\newblock \bibinfo{title}{{Identifying respiratory-related evoked potentials}}.
\newblock \bibinfo{journal}{Engineering in medicine and biology magazine}
  \bibinfo{volume}{14}, \bibinfo{pages}{174--178}.
\bibitem[{MacDonald et~al.(2006)MacDonald, Nyberg and B{\"a}ckman}]{mcdonald}
\bibinfo{author}{MacDonald, S.W.}, \bibinfo{author}{Nyberg, L.},
  \bibinfo{author}{B{\"a}ckman, L.}, \bibinfo{year}{2006}.
\newblock \bibinfo{title}{Intra-individual variability in behavior: links to
  brain structure, neurotransmission and neuronal activity}.
\newblock \bibinfo{journal}{Trends in Neuroscience} \bibinfo{volume}{29},
  \bibinfo{pages}{474--480}.
\bibitem[{Merzagora et~al.(2006)Merzagora, Bunce, Izzetoglu and Onaral}]{anna}
\bibinfo{author}{Merzagora, A.C.}, \bibinfo{author}{Bunce, S.},
  \bibinfo{author}{Izzetoglu, M.}, \bibinfo{author}{Onaral, B.},
  \bibinfo{year}{2006}.
\newblock \bibinfo{title}{{Wavelet analysis for EEG feature extraction in
  deception detection}}, in: \bibinfo{booktitle}{Proceedings of Engineering in
  medicine and biology society EMBS'06 28th Annual Conference of the IEEE},
  \bibinfo{organization}{IEEE}. pp. \bibinfo{pages}{2434--2437}.
\bibitem[{Nielsen et~al.(2006)Nielsen, Kamavuako, Andersen, Lucas and
  Farina}]{niel}
\bibinfo{author}{Nielsen, M.}, \bibinfo{author}{Kamavuako, E.N.},
  \bibinfo{author}{Andersen, M.M.}, \bibinfo{author}{Lucas, M.F.},
  \bibinfo{author}{Farina, D.}, \bibinfo{year}{2006}.
\newblock \bibinfo{title}{{Optimal wavelets for biomedical signal
  compression}}.
\newblock \bibinfo{journal}{Medical and Biological Engineering and Computing}
  \bibinfo{volume}{44}, \bibinfo{pages}{561--568}.
\bibitem[{{Quian Quiroga} et~al.(2001){Quian Quiroga}, Sakowitz, Basar and
  Sch\"{u}rmann}]{quian}
\bibinfo{author}{{Quian Quiroga}, R.}, \bibinfo{author}{Sakowitz, O.W.},
  \bibinfo{author}{Basar, E.}, \bibinfo{author}{Sch\"{u}rmann, M.},
  \bibinfo{year}{2001}.
\newblock \bibinfo{title}{{Wavelet transform in the analysis of the frequency
  composition of evoked potentials}}.
\newblock \bibinfo{journal}{Brain Research Protocols} \bibinfo{volume}{8},
  \bibinfo{pages}{16--24}.
\bibitem[{Rafiee et~al.(2011)Rafiee, Rafiee, Prause and Schoen}]{rafee}
\bibinfo{author}{Rafiee, J.}, \bibinfo{author}{Rafiee, M.A.},
  \bibinfo{author}{Prause, N.}, \bibinfo{author}{Schoen, M.P.},
  \bibinfo{year}{2011}.
\newblock \bibinfo{title}{{Wavelet basis functions in biomedical signal
  processing}}.
\newblock \bibinfo{journal}{Expert Systems with Applications}
  \bibinfo{volume}{38}, \bibinfo{pages}{6190--6201}.
\bibitem[{Rondik and Ciniburk(2011)}]{tomas}
\bibinfo{author}{Rondik, T.}, \bibinfo{author}{Ciniburk, J.},
  \bibinfo{year}{2011}.
\newblock \bibinfo{title}{{Detection of ERP components - comparison of basic
  methods and their modifications}}, in: \bibinfo{booktitle}{Proceedings of 4th
  INCF Congress of Neuroinformatics}.
\bibitem[{Samar et~al.(1999)Samar, Bopardikar, Rao and Swartz}]{vince}
\bibinfo{author}{Samar, V.J.}, \bibinfo{author}{Bopardikar, A.},
  \bibinfo{author}{Rao, R.}, \bibinfo{author}{Swartz, K.},
  \bibinfo{year}{1999}.
\newblock \bibinfo{title}{{Wavelet analysis of neuroelectric waveforms: a
  conceptual tutorial}}.
\newblock \bibinfo{journal}{Brain and Language} \bibinfo{volume}{66},
  \bibinfo{pages}{7--60}.
\bibitem[{Samar et~al.(1992)Samar, Raghuveer, Swartz, Rosenberg and
  Chaiyaboonthanit}]{vince1}
\bibinfo{author}{Samar, V.J.}, \bibinfo{author}{Raghuveer, M.R.},
  \bibinfo{author}{Swartz, K.P.}, \bibinfo{author}{Rosenberg, S.},
  \bibinfo{author}{Chaiyaboonthanit, T.}, \bibinfo{year}{1992}.
\newblock \bibinfo{title}{{Wavelet decomposition of event related potentials:
  toward the definition of biologically natural components}}, in:
  \bibinfo{booktitle}{Proceedings of IEEE Sixth SP Workshop on Statistical
  Signal and Array Processing}, \bibinfo{organization}{IEEE}. pp.
  \bibinfo{pages}{38--41}.
\bibitem[{Samar et~al.(1995)Samar, Swartz and Raghuveer}]{vince3}
\bibinfo{author}{Samar, V.J.}, \bibinfo{author}{Swartz, K.P.},
  \bibinfo{author}{Raghuveer, M.R.}, \bibinfo{year}{1995}.
\newblock \bibinfo{title}{{Multiresolution analysis of event-related potentials
  by wavelet decomposition}}.
\newblock \bibinfo{journal}{Brain and Cognition} \bibinfo{volume}{27},
  \bibinfo{pages}{398--438}.
\bibitem[{Segalowitz et~al.(1997)Segalowitz, Dywan, Unsal et~al.}]{sego}
\bibinfo{author}{Segalowitz, S.}, \bibinfo{author}{Dywan, J.},
  \bibinfo{author}{Unsal, A.}, et~al., \bibinfo{year}{1997}.
\newblock \bibinfo{title}{Attentional factors in response time variability
  after traumatic brain injury: an erp study}.
\newblock \bibinfo{journal}{Journal of International Neuropsychological
  Society} \bibinfo{volume}{3}, \bibinfo{pages}{95--107}.
\bibitem[{Serby et~al.(2005)Serby, Yom-Tov and Inbar}]{serby}
\bibinfo{author}{Serby, H.}, \bibinfo{author}{Yom-Tov, E.},
  \bibinfo{author}{Inbar, G.F.}, \bibinfo{year}{2005}.
\newblock \bibinfo{title}{{An improved P300-based brain-computer interface}}.
\newblock \bibinfo{journal}{IEEE Transactions on Neural Systems and
  Rehabilitation Engineering} \bibinfo{volume}{13}, \bibinfo{pages}{89--98}.
\bibitem[{Thesasiri et~al.(2008)Thesasiri, Charoensuk and
  Sittiprapaporn}]{watun}
\bibinfo{author}{Thesasiri, W.}, \bibinfo{author}{Charoensuk, W.},
  \bibinfo{author}{Sittiprapaporn, W.}, \bibinfo{year}{2008}.
\newblock \bibinfo{title}{{Development of algorithm for erp signal detection}},
  in: \bibinfo{booktitle}{Proceedings of the 3rd International symposium on
  biomedical engineering (ISBME 2008)}, pp. \bibinfo{pages}{175--179}.
\bibitem[{Torrence and Compo(1998)}]{terrence}
\bibinfo{author}{Torrence, C.}, \bibinfo{author}{Compo, G.P.},
  \bibinfo{year}{1998}.
\newblock \bibinfo{title}{A practical guide to wavelet analysis}.
\newblock \bibinfo{journal}{Bulletin of the American Meteorological society}
  \bibinfo{volume}{79}, \bibinfo{pages}{61--78}.
\bibitem[{Trejo and Shensa(1999)}]{trejo}
\bibinfo{author}{Trejo, L.J.}, \bibinfo{author}{Shensa, M.J.},
  \bibinfo{year}{1999}.
\newblock \bibinfo{title}{{Feature extraction of event-related potentials using
  wavelets: an application to human performance monitoring}}.
\newblock \bibinfo{journal}{Brain and Language} \bibinfo{volume}{66},
  \bibinfo{pages}{89--107}.
\bibitem[{Walker(2002)}]{walker}
\bibinfo{author}{Walker, J.S.}, \bibinfo{year}{2002}.
\newblock \bibinfo{title}{{A primer on wavelets and their scientific
  applications}}.
\newblock \bibinfo{publisher}{CRC press}.
\bibitem[{Wilkinson and Seales(1978)}]{robert}
\bibinfo{author}{Wilkinson, R.T.}, \bibinfo{author}{Seales, D.M.},
  \bibinfo{year}{1978}.
\newblock \bibinfo{title}{{EEG event-related potentials and signal detection.}}
\newblock \bibinfo{journal}{Biological Psychology} \bibinfo{volume}{7},
  \bibinfo{pages}{13--28}.

\end{thebibliography}

\end{document}